\documentclass[a4paper,11pt]{article}
\usepackage{jinstpub} 
\usepackage{lineno}

\title{Why would you put a flashlight in a dark matter detector?}

\author[a,b]{R. Gibbons,}\emailAdd{rgibbons@berkeley.edu}
\author[b]{H. Chen,}
\author[b]{S.J. Haselschwardt,}
\author[b]{Q. Xia,} 
\author[b]{and P. Sorensen}\emailAdd{pfsorensen@lbl.gov}
\affiliation[a]{University of California, Berkeley, Department of Physics, Berkeley, CA 94720, USA}
\affiliation[b]{Lawrence Berkeley National Laboratory, 1 Cyclotron Road, Berkeley, CA 94720, USA}

\abstract{
Silicon photomultipliers (SiPMs) are solid-state, single-photon sensitive, pixelated sensors whose usage for scintillation detection has rapidly increased over the past decade. 
It is known that the avalanche process within the device, which renders a single photon detectable, can also generate secondary photons which may be detected by a separate device. 
This effect, known as external crosstalk, could potentially degrade the science goals of future xenon dark matter experiments.
In this article, we measure the effect of external crosstalk in a dual-phase, liquid xenon time projection chamber fully instrumented with SiPMs.
We then consider the implications for a future xenon dark matter experiment utilizing SiPMs and discuss possible solutions.
}

\begin{document}
\maketitle
\flushbottom

\section{Introduction}
\label{sec:intro}
Advances in solid-state technologies have led to increased usage of silicon photomultipliers (SiPMs) for scintillation light detection in particle physics instrumentation~\cite{Simon:2018xzl}. In particular, SiPMs are being actively considered for use in experiments aimed at the direct detection of dark matter such as the proposed XLZD experiment~\cite{Aalbers:2022dzr} and for potential upgrades to the LUX-ZEPLIN (LZ) detector~\cite{Kravitz:2022mby,Chen:2023,haselschwardt2023measurement,Lippincott:2017yst}. 
The appeal is significant compared to photomultiplier tubes (PMTs): more compact footprint in both size and amount of radioimpurities, resilience to magnetic fields, lower operating voltages, and a naturally pixelated photosensitive area which could improve event reconstruction. 
As a brief description, a SiPM is a pixelated array of avalanche photodiodes: p-n junctions reverse-biased beyond their breakdown voltage.
When a pixel detects a primary photon, the resulting Geiger-mode avalanche of charge carriers also emits secondary photons~\cite{McLaughlin:2021xat,2009NIMPA.610...98M}. 
This side effect is generic to silicon avalanche devices~\cite{PhysRev.100.700}.
These secondary photons can themselves be detected by different pixels in the SiPM and therefore create excess, spurious signal, an effect referred to as \emph{optical crosstalk}.\footnote{Other effects exist, notably afterpulsing; these are discussed in e.g., ref.~\cite{Gallina:2022zjs} and are not considered further here.}
Thus, a disadvantage of SiPMs is the intrinsic production of excess signal, in the form of crosstalk, upon photon detection, an effect that scales non-linearly with device gain~\cite{Baudis:2018pdv,Boulay:2022rgb}.

Optical crosstalk can be easily calibrated so long as it is contained internally within the originating device. In this case the effect is typically referred to as \emph{internal crosstalk}.
If multiple SiPMs are instrumented within a detector, crosstalk between different devices can occur.
This is known as \emph{external crosstalk}. Because the secondary photon has escaped the originating device and been detected by another SiPM, calibration is no longer straightforward. 
In this manner, a SiPM is unfortunately behaving as a pulsed flashlight.
Indeed, calibration of external crosstalk is not possible at the single-device level, and can only be measured by other devices within the particle detector system.
While numerous groups have studied internal crosstalk~\cite{Baudis:2018pdv,Gallina:2022zjs,Pershing:2022eka,Wang:2021gof}, limited attention~\cite{Moharana:2022pvg,Boulay:2022rgb,Rebeiro:2022cpad} has been given to studies of external crosstalk between devices within a larger detector.

In general, extraneous photon sources in a scintillation detector (of any type) are undesirable, as mismodelling of detector response may lead to spurious signal excesses.
Even if correctly modeled, we expect the presence of external crosstalk to corrupt the low-energy thresholds of dark matter searches, such as in xenon and argon-based dark matter detectors~\cite{Razeto:2022par}.
This is because a single, thermally-generated avalanche (dark count) could induce a coincident signal in a separate device.
The probability for spurious $N$-fold coincident events at detector threshold thus rises accordingly, making it challenging to distinguish genuine signals at low energy.
For a dark matter detector in which sensitivity improves with decreasing energy threshold, the usage of SiPMs could prove detrimental.

In this article, we have measured the magnitude of external crosstalk in a dual-phase, liquid xenon time projection chamber (TPC) fully instrumented with SiPMs, with the goal of understanding the severity of this effect for future xenon dark matter experiments. We find the probability that a single detected photon generates external crosstalk can grow to $\gtrsim5\%$ in our detector.
We interpolate our measurement to a future xenon dark matter experiment, showing the low-energy threshold would worsen, and comment on potential paths forward.

\section{Measurement of external crosstalk}
\label{sec:exp}

\subsection{Detector description}
\label{subsec:det}

\begin{figure}
    \centering
    \includegraphics[width=.98\textwidth,angle=0]{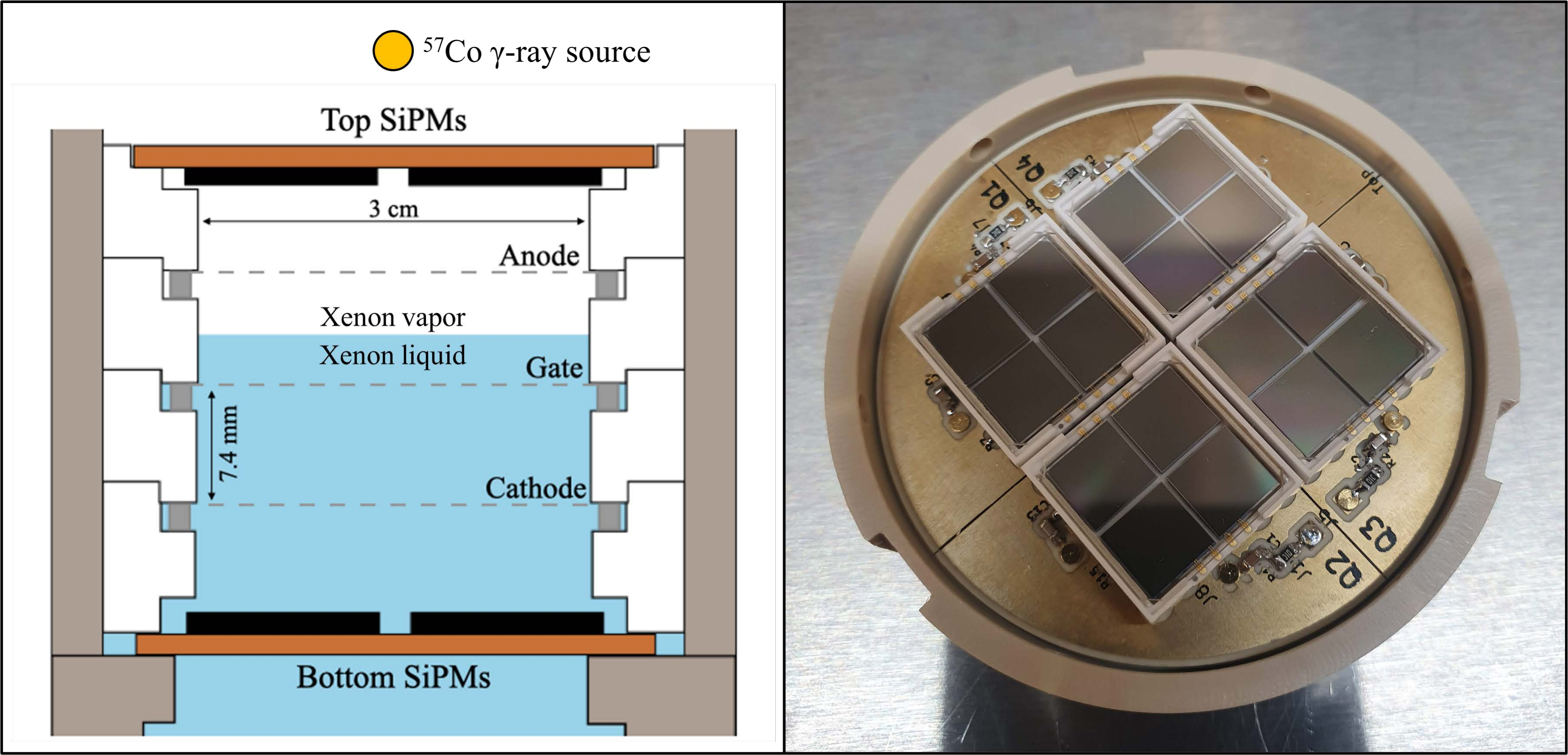}
    \caption{\emph{Left:} cross section schematic of the TPC. 
    Dimensions are approximately to scale, except for the $^{57}$Co source, which is placed $\sim10~$cm above the TPC.
    \emph{Right:} a photo of one of the two SiPM arrays. Each of the four devices (quadrants) contains four separate SiPMs, covered by a single quartz window, for a total of sixteen channels per array.
    }
    \label{fig:diagram}
\end{figure}

Full details of the dual-phase, liquid xenon TPC may be found in~\cite{Kravitz:2022mby}.
As shown in figure~\ref{fig:diagram} (left), a cylindrical TPC is contained with PTFE walls holding three electrodes: cathode, gate, and anode, as well as two SiPM arrays located at either end of the TPC.
This design is meant to emulate an LZ-like detector.
The two SiPM arrays each contain four Hamamatsu VUV4 S13371-6050CQ SiPM devices for a total of 32 (31 operating in this study) SiPM channels.
One of the arrays is pictured in figure~\ref{fig:diagram} (right).
Each array is mounted on a circuit board which provides an independent bias voltage to each device. This bias voltage is low-pass filtered with $R=10~$k$\Omega$ and $C=10~$nF. Additionally, a $50~\Omega$ resistance to ground is located immediately after the output of each device.   
The output signal of each SiPM channel is sent directly to one of three digitizers (CAEN V1730S) without amplification. The digitizer bandwidth is 250 MHz.

An external $^{57}$Co source is present during all measurements, including SiPM calibrations.
The source emits two prominent $\gamma$-rays with energies of 122~keV ($\approx88\%$ branching fraction) and 136~keV ($\approx12\%$ branching fraction) which mainly photoabsorb in the liquid xenon.
These photoabsorption events are seen as a merged peak in energy.
Photoabsorption, or any other scattering event, first creates the prompt scintillation signal, S1.
Second, ionized electrons are drifted in an electric field and are extracted into the vapor, at which point they are subject to a high electric field creating the ionization signal, S2.
The average field strength in the drift region, the region between cathode and gate, is 270~V/cm.
In the extraction region, located between the gate and anode electrodes, an electric field of $\approx 9.7~$kV/cm is maintained in the vapor ($\approx 5.2~$kV/cm in liquid). 
The measured drift time $= t_{S2} - t_{S1}$ determines the vertical coordinate of the interaction vertex, while the reconstructed $(x,y)$ position is calculated with a weighted average of the S2 signal distribution seen in the top SiPM array.
Photoabsorption events are single scatters (SS) which are defined to have only one S1,S2 pair.

Event acquisition is triggered when any two channels exceed a 10~mV threshold within an 8~ns coincidence window. 
This trigger has full efficiency for $^{57}$Co photoabsorption events.
The SiPM signal waveforms are digitized at 500 MS/s using the three digitizers described above. 
The subsequent processing differs from ref.~\cite{Kravitz:2022mby}, but is identical to that in ref.~\cite{haselschwardt2023measurement}.
To remove high-frequency noise, the signal trace from each channel is successively filtered using two boxcar filters with area-preserving kernels and widths of 5 and 21 samples each.
Waveforms are then analyzed by a pulse finding algorithm optimized for selecting SS events, after which pulse variables such as pulse area and width are calculated.
As described in the next section, a different trigger type and pulse finding algorithm is used to acquire and analyze single-photon pulses.

\subsection{SiPM calibration}
\label{subsec:sipm}

We calibrate each SiPM device using single-photon pulses at detector conditions described above with SiPM bias voltages ranging from 49~V to 54~V.
The average breakdown voltage of the SiPMs, shown in figure~\ref{fig:speSweep} (right), is 45.8~V when operating at $\approx 172~$K.
The proceeding voltage range then corresponds to 3.2~V to 8.2~V average overvoltage.
Data acquisition occurs 500~$\mu$s after an S2 pulse occurs, determined by a pulse width trigger on the S2.
At these times, strings of single-photon pulses are present, caused by the detector effects of delayed photon and electron emission.
Using these pulses for single-photon calibration avoids bias from a direct pulse trigger, and ensures a larger average pulse multiplicity than a random trigger.

\begin{figure}[p]
    \centering
    \includegraphics[width=.98\textwidth,angle=0]{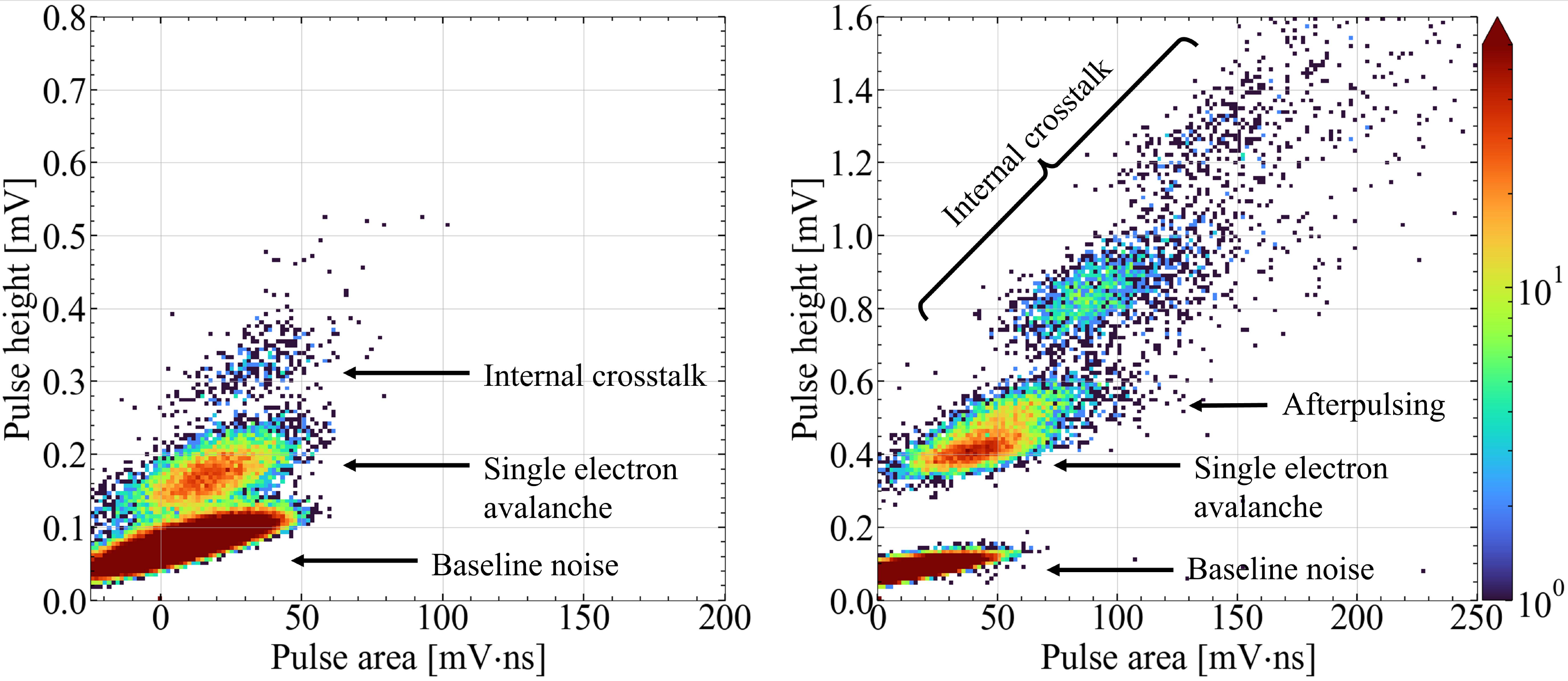}
    \caption{\emph{Left:} the height and area spectrum of pulses induced by single-photons for SiPM channel 0 biased at 49 V.
    The SiPM temperature is $\approx172~$K.
    Pulses larger than the single electron avalanche peak are attributed to internal crosstalk.
    \emph{Right:} the similar spectrum as the left panel for the same channel biased at 54 V.
    Here, internal crosstalk is significantly more pronounced.
    In addition, the effect of afterpulsing is seen as an upper deformation on the single electron avalanche peak.
    We emphasize that external crosstalk cannot be seen with this single-device analysis.
    }
    \label{fig:speHist}
\end{figure}

\begin{figure}[p]
    \centering
    \includegraphics[width=.98\textwidth,angle=0]{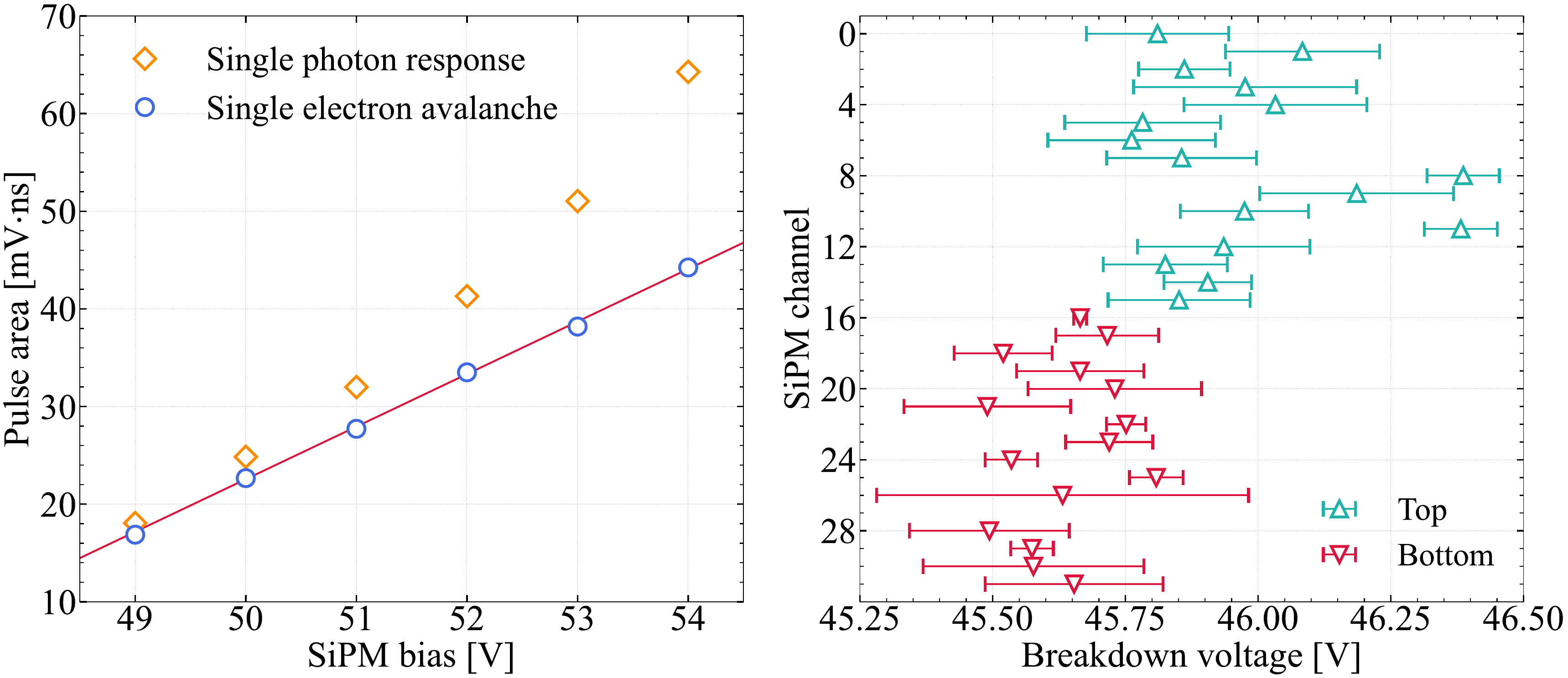}
    \caption{\emph{Left:} The calibration values for SiPM channel 0 across bias voltages.
    Error bars are smaller than the markers.
    A linear fit to the single electron avalanche (blue circles) from 50 V -- 54 V is used to extrapolate the breakdown voltage of each channel.
    The single photon response (orange diamonds) includes internal crosstalk and afterpulsing, as shown in figure \ref{fig:speHist}. 
    The difference between these calibration values is found to be consistent with ref.~\cite{Gallina:2022zjs}.
    \emph{Right:} the breakdown voltage for each SiPM. 
    Error bars are the errors on the linear fit parameters, and we again note SiPM channel 27 was not functioning in this study.
    Variations in breakdown voltages are mainly due to differences in temperature, which can be seen with the warmer top array in the xenon vapor, and the cooler bottom array in the liquid.
    }
    \label{fig:speSweep}
\end{figure}

Pulses are selected in each channel as follows: first the maximum amplitude sample is found, and the pulse bounds are set as a fixed window of [$-$240~ns, $+$560~ns) about this sample.
The width of this window is chosen to be larger than that of both single-photons and $^{57}$Co S1 pulses such that SiPM secondary effects (crosstalk, afterpulsing) are encapsulated.
The timing of the SiPM excess signals has been measured to occur mostly between 10--100ns after the primary avalanche~\cite{Gallina:2022zjs}.
After the first pulse is found, it is removed from the waveform, and the process is repeated up to four more times for a maximum possible total of five pulses per event per channel.
Pulses found to overlap with previously found pulses are excluded from the analysis.
To minimize bias, the amplitude of the final pulse selected (therefore the lowest amplitude pulse) is required to be consistent with baseline noise, otherwise that event is excluded.
In addition, as a data quality measure, waveforms whose total area is above a certain threshold are excluded.
The resulting spectrum of pulses is shown in figure~\ref{fig:speHist}.

SiPM response can be quantified in two ways: (1) the single electron avalanche response, normally referred to as a gain calibration, and (2) the single photon response, which includes the secondary internal crosstalk and afterpulsing effects.
The latter quantity being the more appropriate measurement of device response.
As discussed above, neither of these calibrations can quantify the amount of external crosstalk, which must be measured separately as performed in the next section.
For both calibrations, a selection in pulse height and area is done to remove baseline noise pulses.
Then, for the single electron avalanche calibration, an additional selection in pulse height is performed to isolate the single avalanche response, and the resulting pulse area spectrum is fit to a normal distribution.
The single photon response is determined by calculating the mean of all pulse areas above baseline noise which, to reiterate, encapsulates internal crosstalk and afterpulsing.
Figure~\ref{fig:speSweep} (left) shows the measurements for both calibrations.
We note that because we utilize delayed photon and electron emission, the single photon response (but not the single avalanche response) will be systematically overestimated due to the chance of a single device reading two or more photons simultaneously.
However, due to the convergence of both calibrations at lower SiPM bias, as shown in figure~\ref{fig:speSweep} (left), we consider this effect to be secondary.

\subsection{Analysis and Results}

The following analysis method allows us to robustly quantify the amount of external, optical crosstalk that is created by the bottom SiPM array and viewed by the top array.
Data using the $^{57}$Co source described above is first collected with both arrays at the same bias voltage.
Then, data is taken with only the top array powered, while all bottom array channels are held at 0~V bias.
This process is repeated for the same range of bias voltages as done in the SiPM calibration, 49~V -- 54~V.
The difference in the $^{57}$Co S1 peak mean measured by the top array between the two cases, bottom on and bottom off, determines the amount of external crosstalk.

We utilize the $^{57}$Co peak seen in the liquid xenon extraction region (described in section~\ref{subsec:det}).
Comparable results are observed using events in the drift region; however the extraction region has a higher rate of events which facilitates data acquisition.
Two selections of event quantities are made to isolate the $^{57}$Co peak: drift time to confine the extraction region, and reconstructed radius to reduce the amount of unwanted background events.
In addition, to mitigate any differences between SiPM calibration and data, the S1 area is determined using the same fixed pulse bound used in SiPM calibration.
During data collection, the xenon vapor pressure was kept at ($1.50 \pm 0.02$)~bar and we did not observe any changes to detector response.

\begin{figure}[t]
    \centering
    \includegraphics[width=.98\textwidth,angle=0]{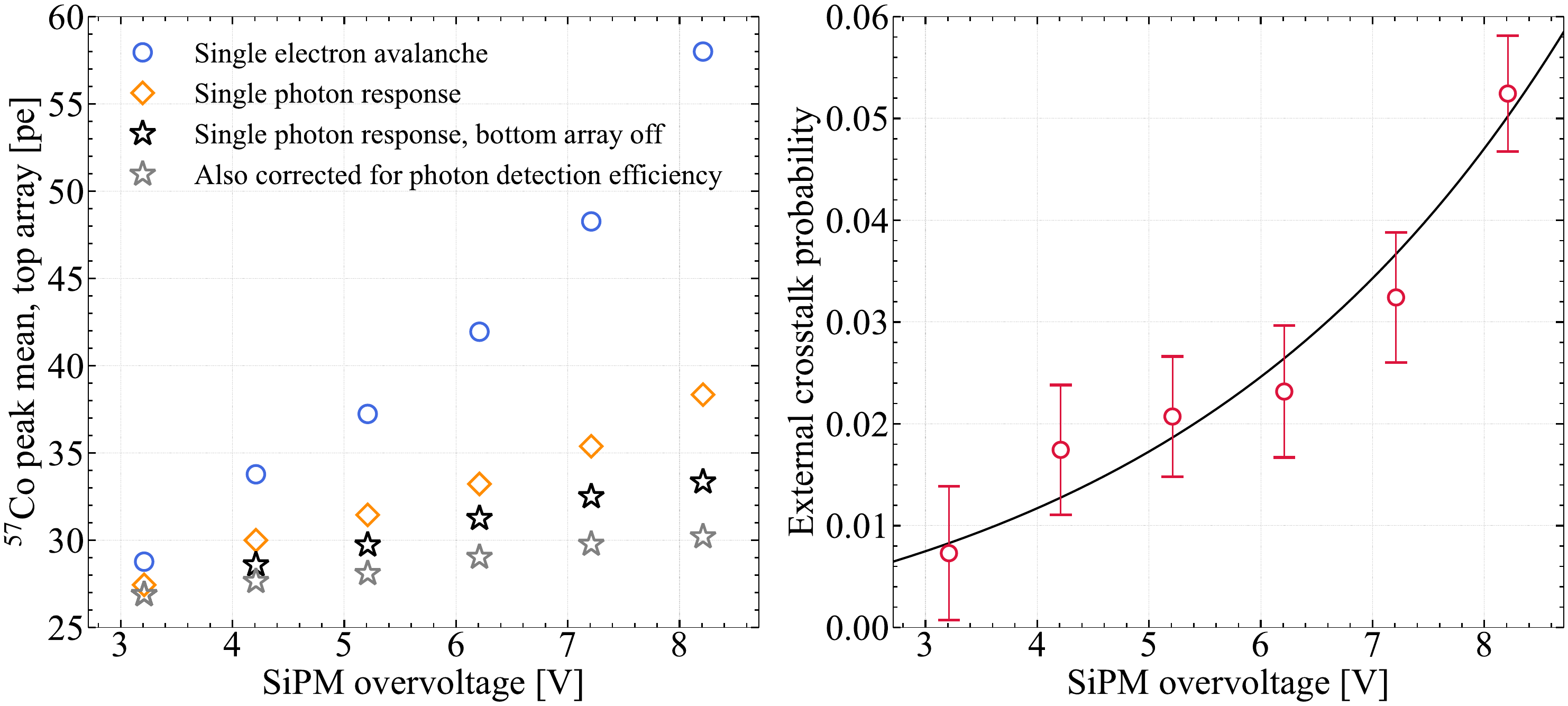}
    \caption{\emph{Left:} the results of the $^{57}$Co peak means seen in the top array for the range of mean SiPM overvoltages.
    Data with both arrays on is calculated using both the single electron avalanche calibration (blue circles) and single photon response calibration (orange diamonds).
    The case where the bottom array is off uses the single photon response calibration (black stars) and is further corrected for differences in photon detection efficiency for xenon VUV scintillation (gray stars).
    Error bars are smaller than the markers.
    \emph{Right:} the probability of generating external crosstalk, calculated by taking the excess signal seen by the top array (the difference between orange diamonds and black stars in the left panel) divided by the signal seen in the bottom array.
    A best fit exponential with the form $y=a(e^{x/b}-1)$ is additionally shown.
    }
    \label{fig:results1}
\end{figure}

The mean number of detected photoelectrons in the top SiPM array due to the $^{57}$Co source is shown in figure~\ref{fig:results1}. We show the response of the top SiPM array, with both arrays on, using the two SiPM calibration values described in section~\ref{subsec:sipm}.
Note that we only use the single photon response calibration in this analysis; the single electron avalanche calibration is simply shown as a reference.
The difference between the bottom-array-on and bottom-array-off data is then a direct measurement of the amount of external crosstalk generated by the bottom array, as seen in the top array. 
The probability of generating external crosstalk from the bottom array and detecting in the top is shown in figure~\ref{fig:results1} (right).
This is calculated by taking the bottom-array-on and bottom-array-off difference, and dividing by the peak seen in the bottom array with both arrays on.
This is an underestimate of the probability, as the bottom array signal will also contain crosstalk originating from the top array.
Most of the S1 light is seen in the bottom array and therefore this correction would be secondary to our calculation.
A final correction is applied to account for the change in VUV photon detection efficiency (PDE) of the devices as a function of bias~\cite{Gallina:2022zjs}.
This is shown in figure~\ref{fig:results1} (left), and is made relative to the lowest SiPM bias used, 49~V.

\section{Discussion}
\label{sec:discussion}
Our expectation is that for a correctly calibrated instrument, the measured number of detected photons from e.g., a $^{57}$Co $\gamma$-ray should be independent of the SiPM bias voltage.
As seen in figure~\ref{fig:results1} (left), correcting for internal crosstalk and afterpulsing, external crosstalk from the bottom array, and accounting for differences in VUV PDE have helped to this end.
Nevertheless, the measured trend is still not flat with SiPM bias voltage. There are likely several reasons for this: (1) external crosstalk generated and detected within the top array and (2) PDE wavelength dependence. 
External crosstalk can be generated and detected by the top array in a similar manner to bottom-to-top external crosstalk, with photons leaving the device, reflecting off of other surfaces, and being detected in the originating array.
More specifically, this can occur through what has been called feedback crosstalk, where exiting photons are reflected off of the SiPM quartz window into another device.
In our arrays, a single quartz window covers four SiPM devices.
Indeed, in refs.~\cite{Moharana:2022pvg,Boulay:2022rgb}, feedback crosstalk is a major component of their measured signal.

With PDE wavelength dependence, we hypothesize this is due to longer wavelengths travelling further in the device before being absorbed~\cite{Green:1995si,PhysRev.99.1151} coupled with the fact an increase in bias voltage may increase the size of the active region of the device (the depletion region).
Therefore, PDE as a function of bias voltage may scale differently for different wavelengths of light.
We have two sources of non-VUV light: external crosstalk~\cite{McLaughlin:2021xat,2009NIMPA.610...98M} and IR scintillation of the xenon~\cite{BRESSI2000254,Piotter:2023qli}, and these may contribute to the persistent increase in signal.
We also considered the effect of the increase in dark count rate for larger bias voltages.
In principle, dark counts could be coincident with the S1 pulses from $^{57}$Co, but this occurs at a rate that is too small to explain the trend measured here~\cite{Baudis:2018pdv}.

One potential systematic in this study is the effect of \emph{electronic} crosstalk between top and bottom SiPM channels through close contact of their signal cables.
However, since the cabling is coaxial at all points where top and bottom signals are in contact, we do not expect this effect to be prominent.
We confirmed this by inspecting waveforms of the bottom array when they are set to 0~V bias, while the top array is above breakdown voltage, and do not see any induced pulses in the bottom array indicative of electronic crosstalk.

\section{Outlook}
\label{sec:outlook}
We have shown that SiPM external crosstalk complicates the device calibration in a bias-voltage-dependent way. 
The measured effect also presents issues for accidental coincidence in scintillators.
To illustrate this, we consider the effect of external crosstalk on the $N$-fold coincidence rate in an arbitrarily sized xenon detector with the same top-bottom array geometry as our detector.
As an example, an $N$-fold coincidence signal can be produced as follows: first, a dark count generated in the bottom array induces a signal via external crosstalk in the top array with a probability $P$.
In turn, this induces yet another signal in the bottom array, again via external crosstalk.
This process can continue on with a single dark count producing signal in $N$ SiPM devices.

We use our measurement of external crosstalk probability in this extrapolation. 
Although we have only measured the bottom-to-top crosstalk probability, we assume the top-to-bottom probability is the same. 
In actuality the two cases would be slightly different due to the presence of the liquid-vapor interface. 
We additionally assume that optical effects concerning the light emitted by the SiPM~\cite{McLaughlin:2021xat}, such as absorption lengths in xenon and surface reflectivity of e.g., PTFE, do not affect the result when scaling to larger detectors.

To quantify the total $N$-fold coincident rate, we first consider the rate of uncorrelated $N$-fold coincident dark counts, without external crosstalk.
That is, $N$ SiPM devices each producing a dark count in a coincident time window.
Given $M$ total SiPM devices, each with dark count rate $R_{DC}$, the rate of $N$-fold uncorrelated coincidence pulses $S_N$ in a coincidence window of length $T$ is
\begin{equation}
    S_N \approx \frac{1}{T} \frac{M!}{(M-N)!} \left[ R_{DC}T\, e^{-R_{DC}T }\right]^{N},
\end{equation}
where $R_{DC}T << 1$.
Then, we consider the contribution of external crosstalk to this rate. 
For $2$-fold coincidences, there is a probability $P$ for a single dark count to produce signal in another device, therefore, the \emph{total} rate of 2-fold coincidences $R_2$ is,
\begin{equation}
    R_2 = S_2 + PS_1.
\end{equation}
Following similar logic for the total rate of 3-fold coincidences $R_3$,
\begin{equation}
    R_3 = S_3 + 2PS_2 + P^2S_1.
\end{equation}
For general $N$-fold coincidence, the total rate is,
\begin{equation}
    R_N = \sum_{i=1}^N \binom{N-1}{i-1}\, P^{N-i} S_i.
\end{equation}

\begin{figure}[p]
    \centering
    \includegraphics[width=.98\textwidth,angle=0]{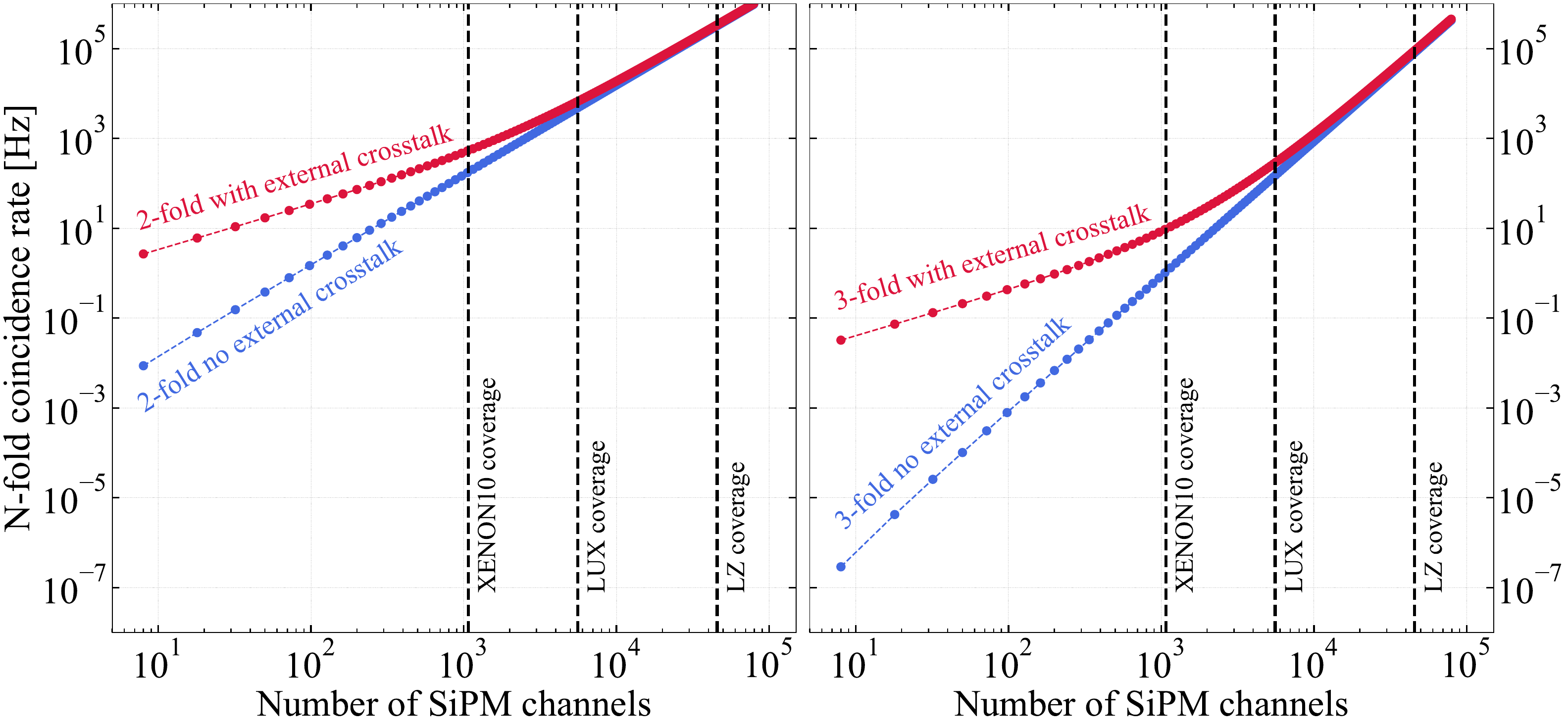}
    \caption{\emph{Left:} the 2-fold coincidence rates for a range of number of SiPM channels using a constant dark count rate as described in the text.
    The rates with external crosstalk correspond to $R_N$, while rates with no external crosstalk correspond to $S_N$.
    For simplicity, we consider each of the two arrays to have the same square number of channels (e.g., 4, 9, 16).
    The vertical dashed lines show the number of SiPMs needed to achieve equal photosensor coverage for the XENON10~\cite{XENON:2010xwm}, LUX~\cite{LUX:2012kmp}, and LZ~\cite{LZ:2019sgr} detectors.
    \emph{Right:} the same as the left panel, but for 3-fold coincidence.
    }
    \label{fig:proj}
\end{figure}

\begin{figure}[p]
    \centering
    \includegraphics[width=.98\textwidth,angle=0]{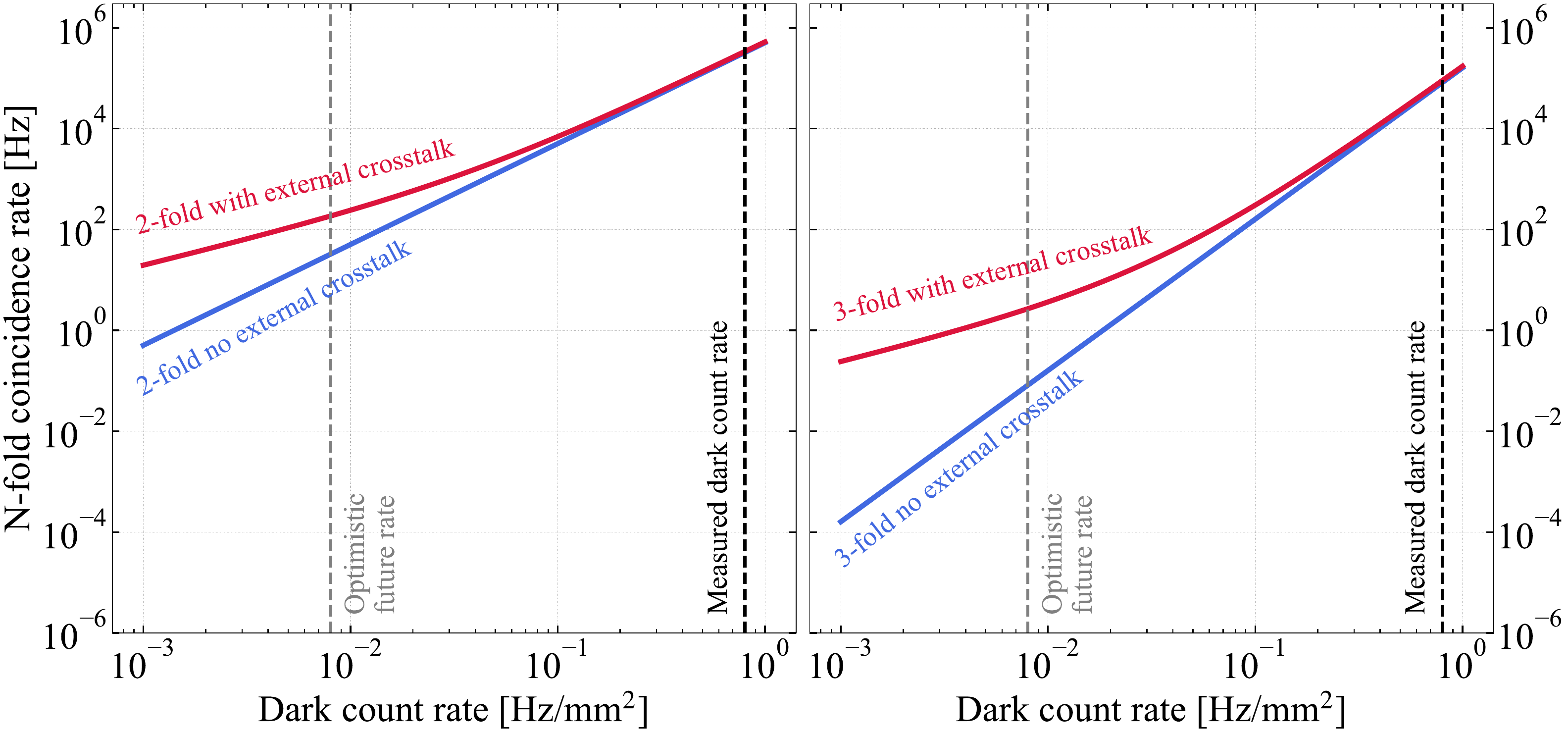}
    \caption{\emph{Left:} the 2-fold coincidence values for a range of dark count rates using a fixed number of SiPM channels equivalent to the photosensor coverage in the LZ detector~\cite{LZ:2019sgr}.
    Two vertical dashed lines are shown: the measured dark count rate from ref.~\cite{Baudis:2018pdv} and a dark count rate two orders of magnitude lower, as suggested in ref.~\cite{DARWIN:2016hyl}.
    \emph{Right:} the same as the left panel, but for 3-fold coincidence.
    }
    \label{fig:proj2}
\end{figure}

For this extrapolation, we use the same SiPM model used in this article.
We assume the top-to-bottom array and bottom-to-top array external optical crosstalk probabilities are both $P=0.012$, based on the value of the best fit curve in figure~\ref{fig:results1} (right) at 4~V overvoltage, which is the manufacturer recommended operating condition.
We consider a fixed dark count rate of $R_{DC} = 0.8~\text{Hz/mm}^2$ as measured at 172~K~\cite{Baudis:2018pdv} and a time coincident window of $T=200~$ns.

The 2-fold and 3-fold coincidence rates, with and without external crosstalk, are shown as a function of the number of SiPM channels in figure~\ref{fig:proj}.
For smaller detectors, we see that the correlated coincident (due to external crosstalk) rate dominates, while for large detectors, the uncorrelated coincident rate (from pile up of dark counts) dominates.
To understand the severity of external crosstalk in such detectors, consider an LZ-sized detector with a 3-fold coincidence threshold on S1, as in refs.~\cite{LUX-ZEPLIN:2022xrq,XENON:2023cxc}.
Using a drift time of $\sim1~$ms~\cite{LUX-ZEPLIN:2022xrq}, one would have a significant probability to have non-zero multiplicity of these pulses, thereby making it impossible to find a real scintillation event of the same magnitude. 
Furthermore, this would significantly increase the number of "accidental" background events: events in which an uncorrelated S1 and S2 pair occur within a physical drift time window and are selected as a physical SS event.
As a comparison, the isolated S1 rate, which directly correlates to the accidental background rate, in LZ is $\mathcal{O}$(1~Hz)~\cite{DQ:cipanp2022}.
Mitigating these backgrounds would require an increase in S1 coincidence threshold of a future experiment, thereby raising the energy threshold.

The uncorrelated coincident rate is already a well-known issue, and it has been estimated a reduction of the dark count rate by two orders of magnitude is required~\cite{DARWIN:2016hyl}. 
We consider then the impact of a reduced dark count rate when including external crosstalk. 
Figure~\ref{fig:proj2} shows both rates $R_N$ and $S_N$ for a fixed number of SiPM devices equivalent to the photosensor coverage in the LZ detector~\cite{LZ:2019sgr}, as a function of dark count rate. 
The presence of external crosstalk causes the $N$-fold coincidence rates to remain significantly larger at lower dark count rates.
Indeed, a reduction of two orders of magnitude from current measured dark count rates would leave the 3-fold coincidence rate at $\mathcal{O}(1~\text{Hz})$.
The necessitates further improvements to reduce external crosstalk in addition to reducing the dark count rate.

We consider two compatible remedies to the effects of external crosstalk for a future xenon dark matter experiment.
First, one may consider the use of optical IR filters with the SiPMs as demonstrated in refs.~\cite{Moharana:2022pvg,Masuda:2021wst}.
The optical crosstalk photons peak in the near infrared, far removed from the mostly VUV scintillation of xenon~\cite{McLaughlin:2021xat}.
This brings several complications: (1) the xenon-filter-SiPM coupling may increase the amount of feedback crosstalk if the indices of refraction are not well-matched~\cite{Masuda:2021wst,Nakamura:2019pwv}. 
(2) a decrease in light collection due to the imperfect transmission of VUV light and the loss of IR scintillation.
(3) an increase in radioactive backgrounds due to additional detector materials present in the xenon.
A second remedy would utilize a hybrid photodetector setup, with SiPMs only used on the top array for more precise $(x,y)$ reconstruction, and PMTs on the bottom array to reduce external crosstalk.
Such a design is already employed in test-stands such as in ref.~\cite{Baudis:2020nwe}.
In this case, the PMTs would not create secondary photons, however, they would still detect some of the secondary photons emitted by the top array, making this an imperfect solution.

\section{Summary}
We have made the first measurement of external crosstalk in a dual-phase liquid xenon TPC, instrumented with a top array and a bottom array of SiPMs. The detector geometry is similar to dark matter search experiments such as LZ. We have shown that SiPM external crosstalk complicates the instrumental calibration by introducing a SiPM bias-voltage dependence. We further find that low-energy scintillation detection threshold may be compromised by accidental coincidences arising from external crosstalk. Ideally, these effects should be remedied at the device level. A future study will examine the effects of feedback external optical crosstalk.

\acknowledgments 
This work is supported by the U.S. Department of Energy, Office of Science,
Office of High Energy Physics, award number DE-AC02-05CH1123.

\bibliographystyle{JHEP}
\bibliography{biblio.bib}

\end{document}